\documentclass[conference, 10pt]{IEEEtran}
\usepackage[left=0.67in,right=0.67in,top=0.7in,bottom=1.1in]{geometry}
\usepackage{amsmath,graphicx,epstopdf}% ,epstopdf,spconf,amssymb,epsfig,fullpage}%, ,graphicx,psfrag,url,amsmath,amsthm,comment}
\usepackage{ulem,color}
\usepackage{epstopdf}
\newcommand{\ie}{\textit{i.e. }}
\newcommand{\eg}{\textit{e.g. }}
\usepackage[utf8]{inputenc}
\usepackage[english]{babel}
\usepackage{multirow}
\usepackage{slashbox}
\usepackage{caption}
\DeclareCaptionLabelFormat{lc}{\MakeLowercase{#1}~#2}
\newcommand\numberthis{\addtocounter{equation}{1}\tag{\theequation}}
\newtheorem{theorem}{Theorem}[section]
\newtheorem{corollary}{Corollary}[theorem]
\newtheorem{lemma}[theorem]{Lemma}
\newtheorem{definition}{Definition}[section]

\begin{document}

\title{Coded Caching with Distributed Storage}

\author{\IEEEauthorblockN{Tianqiong Luo, Vaneet Aggarwal, and Borja Peleato}
\IEEEauthorblockA{Electrical and Computer Engineering\\
Purdue University\\ West Lafayette IN 47907
Email: \{luo133,vaneet,bpeleato\}@purdue.edu}
}

\bibliographystyle{IEEEtran}
\maketitle

\begin{abstract}
Content delivery networks store information distributed across multiple servers, so as to balance the load and avoid unrecoverable losses in case of node or disk failures. Coded caching has been shown to be a useful technique which can reduce peak traffic rates by pre-fetching popular content at the end users and encoding transmissions so that different users can extract different information from the same packet. On one hand, distributed storage limits the capability of combining content from different servers into a single message, causing performance losses in coded caching schemes. But, on the other hand, the inherent redundancy existing in distributed storage systems can be used to improve the performance of those schemes through parallelism.

This paper designs a scheme combining distributed storage of the content in multiple servers and an efficient coded caching algorithm for delivery to the users. This scheme is shown to reduce the peak transmission rate below that of state-of-the-art algorithms.
\end{abstract}

\section{Introduction}
\label{s-intro}
For several decades, CPUs have doubled their speed every two years in what is commonly known as Moore's law, but the storage technology has not been able to keep up with this trend: magnetic hard drives have steadily increased their capacity, but not their speed. Current computers and communication networks are not limited by the speed at which information can be processed, but rather by the speed at which it can be read, moved, and written. Furthermore, the recent information explosion is driving an exponential increase in the demand for data, which is not expected to slow down any time soon. Users and applications demand more data at higher speeds, straining the devices and networks to their maximum capabilities.

The IT industry has addressed this problem through parallelism and caching: instead of using a single high capacity storage drive to serve all the requests, networks usually distribute popular files across multiple independent servers that can operate in parallel and cache part of the information at intermediate or final nodes. %files are split among multiple drives that can operate in parallel for faster storage and recovery. This is known as data striping. Similarly, a single server is often insufficient to fulfill all the requests during peak demand periods, so networks usually have multiple independent servers with separate queues of requests.
This paper proposes and analyzes multiple caching mechanisms for multi-server systems with different system parameters. Previous literature has addressed coded caching for single server systems and distributed storage without caching but, to the extent of our knowledge, this is the first work that considers both coded caching at the users and distributed storage at the servers.
Furthermore, it provides solutions for systems with and without file striping (\ie with files split among multiple servers and with whole files stored in each server).

Distributed storage deals with how the information is stored at the servers. Disk failures are very common in large storage systems, so they need to have some amount of redundancy. Erasure codes have recently sparked a renewed interest from the research community for this task. Files are encoded and distributed among a set of nodes (disks, servers, etc.) in such a way that the system can recover from the failure of a certain number of nodes~\cite{dimakis2010network}, \cite{ernvall2013capacity}. One widely used distributed storage technique based on erasure codes is RAID (redundant array of independent disks). It combines multiple storage nodes (disks, servers, etc.) into a single logical unit with data redundancy. Two of the most common are RAID-4 and RAID-6, consisting of block-level striping with one and two dedicated parity nodes, respectively \cite{corbett2004row, plank2009raid}. Most large scale systems use some form of RAID with striping across multiple storage drives, but store or replicate whole files as a single unit in the network nodes (\eg data centers)~\cite{balasubramanian2014sap}. This increases the peak rate, but it also simplifies book-keeping and deduplication, improves security, and makes the network more flexible.

Coded caching deals with the high temporal variability of network traffic: the peak traffic in the network is reduced by pre-fetching popular content in each receiver's local cache memory during off-peak hours, when resources are abundant. Coded caching has also recently become quite popular among the coding community, starting with the work by Maddah-Ali and Niesen in~\cite{maddah2014fundamental}, which focused on how a set of users with local memories can efficiently receive data from a single server through a common link. Their seminal paper proposed a caching and delivery scheme offering a worst case performance within a constant factor of the information-theoretic optimum, as well as upper and lower bounds on that optimum. The lower bounds were later refined in~\cite{ghasemi2015improved} and new schemes were designed to consider non-uniform file sizes and popularity~\cite{niesen2014coded,zhang2015coded,zhang2015code}; multiple requests per user~\cite{ji2014caching, ji2015caching}; variable number of users~\cite{hachem2015effect}; and multiple servers with access to the whole library of files~\cite{shariatpanahi2015multi}.

Maddah-Ali and Niesen's work in \cite{maddah2014fundamental} caches the information uncoded and encodes the transmitted packets. This scheme performs well when the cache size is relatively large, but a close inspection shows that there are other cases in which its performance is far from optimal. Tian and Chen's recent work in \cite{tian2016caching} designs a new algorithm which encodes both the cached and transmitted segments to achieve a better performance than~\cite{maddah2014fundamental} when the cache size is small or the number of users is greater than the number of files. However, this scheme also focuses on a single server system. In this paper, we aim to design a joint storage and transmission protocol for the multi-server multi-user system.

Summarizing, prior work on distributed storage has studied how a single user can efficiently recover data distributed across a set of nodes and prior work on coded caching has studied how a set of users with local memories can efficiently receive data from a single node. However, to the extent of our knowledge, it has not been studied how the cache placement and content delivery should be performed when multiple nodes send data to multiple users through independent channels. %We aim to design a joint storage and transmission protocol for the multi-server multi-user system in this paper.
In this paper, we aim to design a joint storage and transmission protocol for the multi-server multi-user system. We combine distributed storage with coded caching utilizing parallelism and redundancy to reduce the peak traffic rate. The main contributions of our paper are: (1) a flexible model for multi-server systems where each files can be divided among multiple servers or kept as a single block in one server; (2) an extension of the coded caching algorithms in \cite{maddah2014fundamental} and \cite{tian2016caching} to striping multi-server systems; (3) new caching and delivery schemes with significantly lower peak rates for the case when files are stored as a single unit in a data server.

The rest of the paper is structured as follows: Section \ref{s-background} introduces the system model and two existing coded caching algorithms for single server systems, namely the one proposed by Maddah-Ali and Niesen in~\cite{maddah2014fundamental} and the interference elimination scheme in~\cite{tian2016caching}. Section~\ref{s-striping} extends both algorithms to a multi-server system with file striping, while Sections~\ref{s-ali}~and~\ref{s-interference} consider the case where servers store whole files. Specifically, Section~\ref{s-ali} extends Maddah-Ali and Niesen's scheme, suitable for systems with large cache capacity, and Section~\ref{s-interference} extends the interference elimination scheme, which provides better performance when the cache size is small. Finally, Section~\ref{s-simulation} provides simulations to support and illustrate our algorithms and section \ref{s-conclude} concludes the paper.

\section{Background}\label{s-background}
This section describes the multi-node multi-server model in~\ref{ss-sysmodel} and then reviews two existing coded caching schemes that constitute the basis for our algorithms. Subsection~\ref{ss-basic} summarizes Maddah-Ali and Niesen's coded caching scheme from~\cite{maddah2014fundamental} and subsection~\ref{ss-interference} summarizes Tian and Chen's interference elimination scheme from~\cite{tian2016caching}.

\subsection{System Model}
\label{ss-sysmodel}

We consider a network with $K$ users\footnote{Servers and users can be anything from a single disk to a computer cluster, depending on the application.} and $N$ files stored in $L$ data servers. Some parts of the paper will also include additional parity servers, denoted parity server $P$ when storing the bitwise XOR of the information in the data servers (RAID-4) and parity server $Q$ when storing a different linear combination of the data (RAID-6). The network is assumed to be flexible, in the sense that there is a path from every server to every user \cite{shariatpanahi2015multi}. Each server stores the same number of files with the same size and each user has a cache with capacity for $M$ files. %Except the $N$ basic files stored in the servers, there are some redundant files consisting of repetitions or combinations of other files.
For the sake of simplicity, this paper assumes that all files have identical length and popularity.

The servers are assumed to operate on independent error-free channels, so that two or more servers can transmit messages simultaneously and without interference to the same or different users. A server can broadcast the same message to multiple users without additional cost in terms of bandwidth, but users cannot share the content of their caches with each other. This assumption makes sense in a practical setting since peer-to-peer content sharing is generally illegal. Also, users typically have an asymmetric channel, with large download capacity but limited upload speed.

Similarly, each server can only access the files that it is storing, not those stored on other servers. A server can read multiple segments from its own files and combine them into a single message, but two files stored on different servers cannot be combined into a single message. However, it will be assumed that servers are aware of the content cached by each user and of the content stored in other servers, so that they can coordinate their messages. This can be achieved by exchanging segment IDs through a separate low-capacity control channel or by maintaining a centralized log.

The problem consists of two phases: placement and delivery. During the placement phase, the content is stored in the user's caches. The decisions on where to locate each file, how to compute the parity, and what data to store in each cache are made based on the statistics for each file's popularity, without knowledge of the actual user requests. In our paper, we assume all the files have the same popularity. The delivery phase starts with each user requesting one of the files. All servers are made aware of these requests and proceed to send the necessary messages.

Throughout the paper, we use subindices to represent file indices and superindices to represent segment indices, so $F_i^j$ will represent the j-th segment from file $F_i$. Some parts of the paper will also use different letters to represent files from different servers. For example, $A_i$ to represent the i-th file from server A and $A_i^j$ to represent the j-th segment from file $A_i$. The paper focuses on minimizing the peak rate (or delay), implicitly assuming that different users request different files. Therefore, we will indistinctly refer to users or their requests.

\subsection{Maddah-Ali and Niesen's scheme}
\label{ss-basic}

The coded caching scheme proposed by Maddah-Ali and Niesen in~\cite{maddah2014fundamental} has a single server storing all the files $\{F_1,F_2\ldots,F_N\}$, and users are connected to this server through a shared broadcast link. Their goal is to design caching and delivery schemes so as to minimize the peak load on the link, \ie the total amount of information transferred from the server to the users.
This scheme splits each file $F_i$ into $\binom{K}{t}$ nonoverlapping segments $F_i^j$ of equal size, $j=1,\ldots \binom{K}{t}$, with $t=\frac{KM}{N}$, and caches each segment in a distinct group of $t$ users. In other words, each subset of $t$ users is assigned one segment from each file for all the users to cache\footnote{Parameter $t$ is assumed to be an integer for the sake of symmetry. Otherwise some segments would be cached more often than others, requiring special treatment during the delivery phase and complicating the analysis unnecessarily.}. In the delivery phase the server sends one message to each subset of $t+1$ users, for a total of $\binom{K}{t+1}$ messages.
This caching scheme ensures that, regardless of which files have been requested, each user in a given subset of $t+1$ nodes is missing a segment that all the others have in their cache. The message sent to that subset of nodes consists of the bitwise XOR of all $t+1$ missing segments: a set of users $\mathbf{S}$ requesting files $F_{i_1}, F_{i_2},\ldots, F_{i_{t+1}}$ would receive the message
\begin{equation}
m^{\mathbf{S}}=F_{i_1}^{j_1}\oplus F_{i_2}^{j_2}\oplus\cdots \oplus F_{i_{t+1}}^{j_{t+1}}, \label{e-MNmessage}
\end{equation}
where $j_k$ is the index for the segment cached by all the users in the set except the one requesting $F_{i_k}$. Each user can then cancel out the segments that it already has in its cache to recover the desired segment. In the worst case, \ie when all users request different files, this scheme yields a (normalized by file size) peak rate of
 \begin{align}
R_C(K,t) & = \binom{K}{t+1}/\binom{K}{t}\nonumber \\
& = K(1-M/N)\frac{1}{1+KM/N}. \label{e-RC_binom}
\end{align}
Under some parameter combinations, broadcasting all the missing segments uncoded could require lower rate than $R_C(K,t)$, so the generalized peak rate is

\begin{align}
%R_C(K,t) & = \min\left\{\binom{K}{t+1}\frac{1}{\binom{K}{t}},N-M\right\}\label{e-RC_binom} \\
%K(1-M/N)\min\left\{\frac{1}{1+KM/N},\frac{N}{K}\right\}.\nonumber
\min\left\{R_C(K,t),N-M\right\}\nonumber
\end{align}
but this paper will ignore those pathological cases, assuming that $N$, $M$, and $K$ are such that $R_C(K,t)\leq N-M$. It has been shown that this peak rate is the minimum achievable for some parameter combinations and falls within a constant factor of the information-theoretic optimum for all others~\cite{maddah2014fundamental}\cite{ghasemi2015improved}.

This scheme, henceforth refered to as ``Maddah's scheme" will be the basis for multiple others throughout the paper. It is therefore recommended that the reader has a clear understanding of Maddah's scheme before proceeding.

\subsection{Interference Elimination}
\label{ss-interference}
A close examination of Maddah's algorithm reveals that it has poor performance when the cache is small and $N\leq K$. Thus, a new coded caching scheme based on interference elimination was proposed by Tian and Chen in~\cite{tian2016caching} for the case where the number of users is greater than the number of files. Instead of caching file segments in plain form, they propose that the users cache linear combinations of multiple segments. After formulating the requests, undesired terms are treated as interference that needs to be eliminated to recover the requested segment. The transmitted messages are designed to achieve this using maximum distance separable (MDS) codes \cite{blom1984optimal}\cite{suh2011exact}.

In the placement phase, this scheme also splits each file into $\binom{K}{t}$ non-overlapping segments of equal size and each segment is cached by $t$ users, albeit combined with other segments. Let $F_{i}^{\mathbf{S}}$, where $\mathbf{S}\subseteq \{1,2,\ldots,K\}$ and $\left|\mathbf{S}\right|=t$, denote the file segment from file $F_i$ chosen to be cached by the users in $\mathbf{S}$. In the placement phase user $k$ collects the file segments
\begin{equation}\label{e-segments_interf}
\{F_{i}^{\mathbf{S}}| i\in \{1,2,\ldots,N\},k\in\mathbf{S}\},
\end{equation}
($P=\binom{K-1}{t-1}N$ in total), encodes them with a MDS code $\mathcal{C}(P_0,P)$ of length $P_0=2\binom{K-1}{t-1}N-\binom{K-2}{t-1}(N-1)$, and stores the $P_0-P$ parity symbols in its cache.

The delivery phase proceeds as if all the files are requested. When only some files are requested, the scheme replaces some users' requests to the ``unrequested files" and proceeds as if all files were requested. A total of $\binom{K-1}{t}$ messages are transmitted (either uncoded or coded) for each file $F_i$, regardless of the requests. Uncoded messages provide the segments that were not cached by the users requesting $F_i$, while coded messages combining multiple segments from $F_i$ are used to eliminate the interference in their cached segments. Each user gathers $\binom{K-2}{t-1}(N-1)$ useful messages which, together with the $P-P_0$ components stored in its cache, are enough to recover all $P$ components in the $\mathcal{C}(P_0,P)$ MDS code. A more detailed description of the messages can be found in~\cite{tian2016caching}.

Therefore, the total number of messages transmitted from the server is $N\binom{K-1}{t}$. In this interference elimination scheme, the following normalized $(M,R)$ pairs are achievable:
\begin{equation}\label{e-eliminationrate}
\left(\frac{t\left[(N-1)t+K-N\right]}{K(K-1)},\frac{N(K-t)}{K}\right),\ t=0,1,\ldots,K.
\end{equation}
This scheme is shown to improve the inner bound given in~\cite{maddah2014fundamental} for the case $N\leq K$ and has a better performance than the algorithm in subsection \ref{ss-basic} when the cache capacity is small.

\subsection{Extension to multiple servers}\label{ss-multiple}

Both of the previous schemes assume that a single server stores all the files and can combine any two segments into a message. Then, they design a list of messages to be broadcast by the server, based on the users' requests. In practice, however, it is often the case that content delivery networks have multiple servers and throughput is limited by the highest load on any one server rather than by the total traffic in the link between servers and users. Shariatpanahi et al. addressed this case in~\cite{shariatpanahi2015multi}, but still assumed that all servers had access to all the files and could therefore compose any message. They proposed a load balancing scheme distributing the same list of messages among all the servers, scaling the peak rate by the number of servers.

If each server only has access to some of the files, the problem is significantly more complicated. The general case, where each segment can be stored by multiple servers and users, is known as the index coding problem. This is one of the core problems of network information theory but it remains open despite significant efforts from the research community \cite{el2010index,bar2011index,chaudhry2008efficient}. Instead of addressing the index coding problem in its general form, we focus on the case where each data segment is stored in a single server, all caches have the same capacity, and users request a single file.

A simple way to generalize the previous schemes to our scenario is to follow the same list of messages, combining transmissions from multiple servers to compose each of them. Instead of receiving a single message with all the segments as shown in Eq.~(\ref{e-MNmessage}), each node would receive multiple messages from different servers. The peak rate for any one server would then be the same as in a single server system.

With parity servers storing linear combinations of the data, the peak rate can be reduced. In general, distributed storage systems use MDS codes for the parity\footnote{Some systems use repetition or pyramid codes \cite{el2010fractional}\cite{yu2014irregular}\cite{huang2013pyramid} to reduce the recovery bandwidth, but this paper will focus on MDS codes.}, so any subset of $L$ servers can be used to generate any message. Therefore, a simple balancing of the load by rotating among all subsets of $L$ servers would scale the peak rate by $\frac{L}{L+L'}$, where $L'$ is the number of parity servers. However, we intend to design caching and delivery algorithms capable of further reducing the peak rate of any one server.

\section{File striping}
\label{s-striping}
The simplest way to extend single-server coded caching algorithms to a multi-server system is to spread each file across all servers. That way, each user will request an equal amount of information from each server, balancing the load. This is called data striping \cite{santos2000comparing} and it is common practice in data centers and solid state drives (SSD), where multiple drives or memory blocks can be written or read in parallel. The users then allocate an equal portion of their cache to each server and the delivery is structured as $L$ independent single-server demands. We now proceed to give a detailed description of how striping can reduce the peak rate of Maddah's scheme, but the same idea can be applied to any other scheme.

Each of the $N$ files $\{F_1,F_2\ldots,F_N\}$ is split into $L$ blocks to be stored in different servers and each block is divided into $\binom{K}{t}$ segments. These segments are denoted by $F_{i}^{(j,m)}$, where $i=1,2,\ldots,N$ represents the file number; $j=1,2,\ldots,\binom{K}{t}$ the segment number; and $m=1,2,\ldots,L$ the block number. The $m$-th server is designed to store the $m$-th segment of each file, that is $F_{i}^{(j,m)}$ for every $i$ and $j$.

The placement is the same as in Maddah's scheme. Each segment is cached by $t$ users, with $\{F_{i}^{(j,1)},F_i^{(j,2)},\ldots,F_i^{(j,L)}\}$ being cached by the same user. We notice that each message transmitted by Maddah's scheme in Eq.~(\ref{e-MNmessage}) can be split into $L$ components
\begin{equation}
F_{i_1}^{(j_1,m)}\oplus F_{i_2}^{(j_2,m)}\oplus\cdots \oplus F_{i_{t+1}}^{(j_{t+1},m)}, \label{e-StripingMessage}
\end{equation}
$m=1,2,\ldots,L$ to be sent by different servers.
%Each component can be taken care by one data server, thus the transmission rate is reduced to $\frac{1}{L}$ of the traditional scheme.
Then the problem can be decomposed into $L$ independent single-server subproblems with reduced file sizes of $\frac{F}{L}$ bits. The subproblems have the same number of users, files, and cache capacity (relative to the file size) as the global problem. %The difference lies in the fact that, regardless of which file a user requests, the load is always evenly distributed among the servers. Users request one segment from each server.
Since all servers can transmit simultaneously, the peak load is reduced to $\frac{1}{L}$ of that in Eq.~(\ref{e-RC_binom}) (Maddah's single server scheme).

If one additional parity server $P$ is available (RAID-4), it will store the bitwise XOR of the blocks for each file, \ie $F_i^{(j,1)}\oplus F_i^{(j,2)}\oplus \cdots \oplus F_i^{(j,L)}$ for all $i$ and $j$. %$i=1,2,\ldots,N$ and $j=1,2,\ldots,\binom{K}{t}$.
Then, server $P$ can take over some of the transmissions, reducing the peak load to $\frac{1}{L+1}$ of that with Maddah's scheme\footnote{The number of segments must be a multiple of $L$ to achieve this reduction, but it is always possible to divide each segment into multiple chunks to fulfil this condition.}. Specifically, instead of having all data servers transmit their corresponding component in Eq.~(\ref{e-StripingMessage}), server $P$ can transmit the XOR of all the components, relieving one data server from transmitting. %For example, if data server $1,2,\ldots,L-1$ and P server is involved in the transmission, then P server just transmits all the exclusive sum of all its corresponding stored segments as shown in Table \ref{t-files}.
The users can combine the rest of the components with this XOR to obtain the missing one. % of each message, that is $F_{i_1j_1L}\oplus F_{i_2j_2L}\oplus\ldots \oplus F_{i_{t+1}j_{t+1}L}$, by cancelling all the $L-1$ components from the message P transmits. Therefore, any L servers out of this L+1 RAID-4 server system are enough to compose the message in one transmission.
Similarly, if two additional parity servers $P$ and $Q$ are available (RAID-6), it is possible to choose any $L$ out of the $L+2$ servers to take care of each set of messages in Eq.~(\ref{e-StripingMessage}), thereby reducing the peak rate to $\frac{1}{L+2}$ of that with Maddah's scheme.

A similar process with identical file splitting can be followed for the interference cancelling scheme, achieving the same scaling of the peak rate: $\frac{1}{L}$ when there is no parity, $\frac{1}{L+1}$ with a single parity server, and $\frac{1}{L+2}$ with two parity servers.

In practice, however, it is often preferred to avoid striping and store whole files as a single unit in each server to simplify the book-keeping, ensure security, and make the network more flexible. The rest of the paper will focus on the case where nodes store entire files, and each user requests a file stored in a specific node.

\section{Scheme 1: Large cache}\label{s-ali}

In this section, we extend Maddah-Ali and Niesen's scheme to the multiple server system. Instead of spreading each file across multiple servers as in Section~\ref{s-striping}, each file is stored as a single unit in a data server, as shown in Table~\ref{t-servers_general}. %We assume there are $L$ data servers $A,B,\ldots,L$ and each server stores $r$ files. The files from different server is marked by the server, for example, files from server A are $A_1,A_2,\ldots,A_r$. The $j$-th segment of file $A_i$ is written as $A_i^j$.

\begin{table}
\centering
\begin{tabular}[h]{|c|c|c|c|}%
\hline
Server A & Server B & $\cdots$ & Server L\\
\hline
$A_1$ & $B_1$ & $\cdots$ & $L_1$ \\
$A_2$ & $B_2$ & $\cdots$ & $L_2$ \\
\vdots & \vdots &  & \vdots \\
$A_r$ & $B_r$ & $\cdots$ & $L_r$ \\
\hline
\end{tabular}
\caption{Files stored in each server}\label{t-servers_general}
\end{table}

The performance of Maddah's scheme in Eq.~(\ref{e-RC_binom}) is highly dependent on the cache capacity $M$. Compared with the interference elimination in section \ref{ss-interference}, the advantage of Maddah's scheme lies in that file segments are stored in plain form instead of encoded as linear combinations. This saves some segments from being transmitted in the delivery phase, but it requires larger cache capacities to obtain coded caching gains. Hence, Maddah's scheme is appropriate when the cache capacity is large.

The placement phase of our algorithm is identical to that in the traditional scheme. For example, in a system with $K=6$ users with cache capacity $M=4$ and $N=8$ files, each file is divided into $20$ segments and each segment is stored by $t=3$ users. Table~\ref{t-caches} indicates the indices of the $10$ segments that each user stores, assumed to be the same for all files without loss of generality.

In order to simplify later derivations, the notation is clarified here. Since the peak rate for the storage system is considered, we assume that all users request different files, hence each user can be represented by the file that it has requested. Denote $\mathbf{S}$ to be the user set and $m_A^{\mathbf{S}}$ to represent the message sent from server $A$ to all the users in $\mathbf{S}$. Furthermore, if $\pmb{\alpha}=\{\alpha_1,\alpha_2,\ldots,\alpha_i\}$ represents a vector of file indices and $\pmb{\gamma}=\{\gamma_1,\gamma_2,\ldots,\gamma_i\}$ represents a vector of segment indices, then $\mathbf{A}_{\pmb{\alpha}}$ represents the set of requests (or users)
\begin{equation}\nonumber
\mathbf{A}_{\pmb{\alpha}}=\{A_{\alpha_1}, A_{\alpha_2}, \ldots, A_{\alpha_i}\}
\end{equation}
and $\mathbf{A}_{\pmb{\alpha}}^{\pmb{\gamma}}$ represents the message
\begin{equation}\nonumber
\mathbf{A}_{\pmb{\alpha}}^{\pmb{\gamma}}=A_{\alpha_1}^{\gamma_1}\oplus A_{\alpha_2}^{\gamma_2}\oplus\ldots \oplus A_{\alpha_i}^{\gamma_i},
\end{equation}
where $A_i^j$ represents the $j$-th segment from the $i$-th file in server $A$.
Similarly, $\mathbf{A}_{\pmb{\alpha}}^{\pmb{\gamma}}\oplus \mathbf{B}_{\pmb{\alpha}}^{\pmb{\gamma}}$ represents the the message:
\begin{equation}\nonumber
\mathbf{A}_{\pmb{\alpha}}^{\pmb{\gamma}}\oplus \mathbf{B}_{\pmb{\alpha}}^{\pmb{\gamma}}=(A_{\alpha_1}^{\gamma_1}\oplus B_{\alpha_1}^{\gamma_1})\oplus\ldots\oplus(A_{\alpha_i}^{\gamma_i}\oplus B_{\alpha_i}^{\gamma_i}).
\end{equation}

 We first explore the multi-server system without parity servers in subsection~\ref{ss-MoreFilesNoParity}. Then we study a simple system with two data and one parity server in subsection~\ref{ss-algorithm}. Finally, we study the cases with one and two parity servers in subsections~\ref{ss-beyond} and \ref{s-raid6}, respectively.
\subsection{No parity servers}
\label{ss-MoreFilesNoParity}

In a system without redundancy, such as the one shown in Table~\ref{t-servers_general}, the servers cannot collaborate with each other. During the delivery phase, each user is assigned to the server storing the file that it requested, and then each data server transmits enough messages to fulfil its requests. Specifically, following Maddah's scheme, a server receiving $m$ requests would need to transmit $\binom{K}{t+1}-\binom{K-m}{t+1}$ messages, \ie one for each group of $t$ users containing at least one of its requesters. The normalized peak rate  for that server would therefore be
\begin{equation}\nonumber
\left(\binom{K}{t+1}-\binom{K-m}{t+1}\right)\left/\binom{K}{t}\right.
\end{equation}
The worst case occurs when all users request files from the same server, \ie $m=K$. Then the peak transmission rate is the same as in the single server system.

\subsection{One parity and two data servers}
\label{ss-algorithm}

This section focuses on a very simple storage system with two data servers and a third server storing their bitwise XOR, as shown in Table~\ref{t-servers}. %The three-server coded caching algorithm proposed is shown to reduce the rate greatly compared to the traditional coded caching algorithm.
Despite each server can only access its own files, the configuration in Table~\ref{t-servers} allows composing any message by combining messages from any two servers. Intuitively, if server $A$ (or $B$) finish its transmission task before the other one, it can work with the parity server to help server $B$ (or $A$). This collaborative scheme allows serving two requests for files stored in the same server in parallel, balancing the load and reducing the worst case peak rate below that achieved without the parity server (see Section~\ref{ss-MoreFilesNoParity}).

However, there is a better transmission scheme where messages from all three servers are combined to get more information across to the users. The basic idea is to include some unrequested segments, as well as the requested ones, in each message from a data server. If the additional segments are well chosen, they can be combined with messages from the parity server to obtain desired file segments. The algorithm developed in this section is based on this idea.

\begin{table}
	\centering
	\begin{tabular}[h]{|c|c|c|}%
		\hline
		Server $A$ & Server $B$ & Server $P$\\
		\hline
		$A_1$ & $B_1$ & $A_1 \oplus B_1$ \\
		$A_2$ & $B_2$ & $A_2 \oplus B_2$ \\
		$\vdots$ & $\vdots$ & $\vdots$ \\
		$A_r$ & $B_r$ & $A_r \oplus B_r$ \\
		\hline
	\end{tabular}
	\caption{Files stored in each server}\label{t-servers}
\end{table}

Just like in Maddah's scheme, data servers will send each message to a set of $t+1$ users and the message will contain the XOR of $t+1$ segments (one for each user). These segments are chosen so that all users except the intended receiver can cancel them out. If the user had requested a file stored by the sender, the message will contain the corresponding segment; otherwise the message will include its complement in terms of the parity in server $P$, {\it i.e.} $A_i^j$ instead of $B_i^j$ and vice versa. Therefore, the contents of each message from server $A$ or $B$ are uniquely determined by the sender and the set of receivers, denoted by $S_1$ or $S_2$ respectively. In the example shown in Table~\ref{t-caches}, the message from server $A$ to $S_1=\{A_1, A_2, A_3, B_4\}$, corresponding to users 1 through 4, will be $m_A^{\mathbf{S_1}}=A_1^{11}\oplus A_2^5\oplus A_3^2\oplus A_4^1$.

\begin{lemma}\label{l-combine}
Let the receivers for servers A and B be
\begin{equation}\nonumber
S_1=\{\mathbf{A}_{\pmb{\alpha}},\mathbf{B}_{\pmb{\beta}},\mathbf{A}_\ast\}\qquad S_2=\{\mathbf{A}_{\pmb{\alpha}},\mathbf{B}_{\pmb{\beta}},\mathbf{B}_\ast\},
\end{equation}
respectively, where $\alpha$ and $\beta$ denote (possibly empty) sets of indices, the $\ast$ denote arbitrary sets, and $S_1\neq S_2$. The corresponding messages are
\begin{equation}\nonumber
m_A^{\mathbf{S_1}}= \mathbf{A}_{\pmb{\alpha}}^{\ast}\oplus\mathbf{A}_{\pmb{\beta}}^{\pmb{\gamma}}\oplus\mathbf{A}_\ast^\ast\qquad m_B^{\mathbf{S_2}}=\mathbf{B}_{\pmb{\alpha}}^{\pmb{\eta}}\oplus\mathbf{B}_{\pmb{\beta}}^{\ast}\oplus\mathbf{B}_\ast^\ast,
\end{equation}
with segment indices chosen so that each user can cancel all but one of the components.
This provides users $\mathbf{B}_{\pmb{\beta}}$ and $\mathbf{A}_{\pmb{\alpha}}$ with some unrequested segments $\mathbf{A}_{\pmb{\beta}}^{\pmb{\gamma}}$ and $\mathbf{B}_{\pmb{\alpha}}^{\pmb{\eta}}$, respectively.  Then server $P$ can send the message
\begin{equation}\nonumber
m_P^{\mathbf{S_1\cap S_2}}=(\mathbf{A}_{\pmb{\alpha}}^{\pmb{\eta}}\oplus \mathbf{B}_{\pmb{\alpha}}^{\pmb{\eta}})\oplus(\mathbf{A}_{\pmb{\beta}}^{\pmb{\gamma}}\oplus \mathbf{B}_{\pmb{\beta}}^{\pmb{\gamma}}),
\end{equation}
to $S_1\cap S_2$, so that each user in $S_1$ and $S_2$ obtains a missing segment and those in the intersection obtain two. These three transmissions are equivalent to messages $m^{\mathbf{S_1}}$ and $m^{\mathbf{S_2}}$ as defined in  Eq.~(\ref{e-MNmessage}) for Maddah's single server scheme. They both provide the same requested segments to their destinations.
\end{lemma}
\begin{IEEEproof}
All the users in $S_1$ and $S_2$ get at least one desired segment, from the server storing their requested file. Those in $S_1\cap S_2$ also receive an unrequested segment from server $A$ or $B$. It only remains to prove that users in $S_1\cap S_2$ can use this unrequested segment to obtain its complement from $m_P^{\mathbf{S_1\cap S_2}}$.

Without loss of generality, consider user $B_{\beta_i}\in S_1\cap S_2$. %, which receives $A_{\beta_i}^{\gamma_i}$ from $m_AmmathbfS_1$ and it should be able to use $m_PmmathbfS_1cap S_2$ to obtain $B_{\beta_i}^{\gamma_i}$.
The set of segment indices $\gamma$ in $m_A^{\mathbf{S_1}}$ were chosen so that user $B_{\beta_i}$ is caching all the segments except the $\gamma_i$-th. Similarly, the set of indices $\eta$ in $m_B^{\mathbf{S_2}}$ was chosen so that $B_{\beta_i}$ is caching all of them (for all files). Therefore, $B_{\beta_i}$ can obtain $A_{\beta_i}^{\gamma_i}$ from $m_A^{\mathbf{S_1}}$ and should be able to cancel all terms from $m_P^{\mathbf{S_1\cap S_2}}$ except $A_{\beta_i}^{\gamma_i}\oplus B_{\beta_i}^{\gamma_i}$. Combining both of these yields the desired segment $B_{\beta_i}^{\gamma_i}$. As long as $S_1\neq S_2$, this segment will be different from the one that $B_{\beta_i}$ obtains from $m_B^{\mathbf{S_2}}$ because there is a one-to-one relationship between segment indices and user subsets.
\end{IEEEproof}

 Take the case in Table \ref{t-caches} as an example. Lemma.~\ref{l-combine} states that if $S_1=\{A_1, A_2, A_3, B_4\}$ and $S_2=\{A_1, A_2, B_1, B_4\}$, we construct $m_A^{\mathbf{S_1}}$, $m_B^{\mathbf{S_2}}$, $m_P^{\mathbf{S_1\cap S_2}}$ as:
\begin{align}
m_A^{\mathbf{S_1}}&=A_1^{11}\oplus A_2^5\oplus A_3^2\oplus A_4^1,\nonumber\\
m_B^{\mathbf{S_2}}&=B_1^{14}\oplus B_2^8\oplus B_1^2\oplus B_4^3,\nonumber\\
m_P^{\mathbf{S_1\cap S_2}}&=(A_1^{14}\oplus B_1^{14})\oplus(A_2^8\oplus B_2^8)\oplus (A_4^1\oplus B_4^1).\nonumber
\end{align}
It is easy to verify that these messages are equivalent to two transmissions in Maddah's scheme, specifically those intended for users $\{A_1, A_2, A_3, B_4\}$ and $\{A_1, A_2, B_1, B_4\}$.
\begin{table}
\centering\small
\begin{tabular}[htp]{|c|cccccc|}%
\hline
Segment$\setminus$ User  &  1 &  2 &  3 & 4 & 5 & 6\\
\hline
1& X&X &X & & & \\
2 &X & X& & X& & \\
3& X& X& & & X& \\
4& X& X& & & &X \\
5& X& &X &X & & \\
6& X& &X & &X & \\
7& X& & X& & &X \\
8& X& & & X&X & \\
9& X& & & X& &X \\
10& X& & & & X&X \\
11& & X&X &X & & \\
12& & X& X& & X& \\
13& & X& X& & & X\\
14& & X& &X &X & \\
15& & X& & X& & X\\
16& & X& & &X &X \\
17& & & X& X& X& \\
18& & & X& X & &X \\
19& & & X& &X &X \\
20& & & &X &X &X \\
\hline
Request& $A_1$ &$A_2$ &$A_3$ &$B_4$& $B_1$&$B_2$\normalsize\\
\hline
\end{tabular}
\caption{Mapping of file segments to user caches. Each cache stores the same 10 segments for every file, marked with X in the table.}\label{t-caches}
\end{table}

\begin{corollary}\label{c-combine2}
Assume $S_1=\{\mathbf{A}_\ast,\mathbf{B}_{\pmb{\beta}}\}$ and $S_2=\{\mathbf{B}_\ast\}$, \ie it only contains requests for server $B$. Then server $P$ sends $m_P^{\mathbf{B}_{\pmb{\beta}}}=\mathbf{A}_{\pmb{\beta}}^{\pmb{\gamma}}\oplus \mathbf{B}_{\pmb{\beta}}^{\pmb{\gamma}}$ to all the users in $\mathbf{B}_{\pmb{\beta}}$ in Lemma~\ref{l-combine}, so that all the users in $S_1$ and $S_2$ get the same segments as in Maddah's scheme. The same holds switching the roles of $A$ and $B$.
\end{corollary}

\begin{IEEEproof}
This is a particular case of Lemma~\ref{l-combine} when $\pmb{\alpha}$ is empty ($\pmb{\beta}$ can be empty or non-empty). %Vice versa, when $\pmb{\beta}$ is empty, the same property also holds.
\end{IEEEproof}

\begin{definition}
If user subsets $S_1$ and $S_2$ fulfill the conditions in Lemma~\ref{l-combine}, we call $(S_1,S_2)$ an \textbf{effective pair}.
\end{definition}

Our goal is to design a scheme equivalent to Maddah's scheme while minimizing the maximum number of messages sent by any server. If two user subsets form an effective pair, the corresponding messages in Maddah's scheme (see Eq.~(\ref{e-MNmessage})) can be replaced by a single transmission from each server. Hence, we wish to make as many effective pairs as possible.

\begin{lemma}\label{l-unpaired}
The peak rate is $\left(\frac{1}{2}+\frac{1}{6}\Delta\right)R_C(K,t)$ for the server system in Table~\ref{t-servers}, where $\Delta$ represents the ratio of unpaired messages and $t=\frac{KM}{N}$.
\end{lemma}
\begin{IEEEproof}
For each effective pair, we can use a single transmission from each server to deliver the same information as two transmissions in Maddah's single server scheme. This contributes $\frac{1}{2}(1-\Delta)R_C(K,t)$ to the total rate. Unpaired messages are transmitted as described in section~\ref{ss-multiple}, that is combining messages from any two out of the three servers. Assuming that this load is balanced among all three servers, the contribution to the total rate is $\frac{2}{3}\Delta R_C(K,t)$. Adding both contributions yields the rate above.
\end{IEEEproof}

The following lemma characterizes the ratio of unpaired user subsets $\Delta$ in the case with symmetric requests (both servers receive the same number of requests).

\begin{lemma}\label{l-pairing}
If the requests are symmetric, then $\Delta=0$ when $t$ is even and $\Delta\leq\frac{1}{3}$ when $t$ is odd. That is, the following peak rate is achievable in the case with symmetric requests:
\begin{displaymath}\numberthis \label{e-rate}
R_T(K,t) = \left\{ \begin{array}{ll}
\frac{1}{2}R_C(K,t)&\textrm{if $t$ is even}\\
&\\
\left(\frac{1}{2}+\frac{1}{6}\Delta\right)R_C(K,t)&\textrm{if $t$ is odd,}
\end{array} \right.
\end{displaymath}
where $R_C(K,t)$ is defined in Eq.~(\ref{e-RC_binom}).
\end{lemma}
\begin{IEEEproof}
A pairing algorithm with these characteristics is presented in the Appendix.
\end{IEEEproof}

Although $\Delta$ can reach $\frac{1}{3}$, in most cases the pairing algorithm in the Appendix performs much better. As an example, Table~\ref{t-caches} has each segment cached by $t=\frac{KM}{N}=3$ users and the normalized peak rate with the pairing algorithm is $\frac{2}{5}$, significantly lower than the $\frac{3}{4}$ with Maddah's single server scheme.

Finally, we are ready to derive an achievable peak rate for a general set of requests, based on the following lemma.
\begin{lemma}\label{l-combine3}
If $(S_1,S_2)$ form an effective pair, then $S'_1=\{S_1,\mathbf{A}_{\pmb{\alpha}}\}$ and $S'_2=\{S_2,\mathbf{A}_{\pmb{\alpha}}\}$ also form an effective pair of a larger dimension. The same holds when an all-B file set is appended instead of the all-A file set $\mathbf{A}_{\pmb{\alpha}}$.
\end{lemma}
\begin{IEEEproof}
The proof is straightforward by observing that $(S'_1,S'_2)$ still fulfills the conditions in Lemma~\ref{l-combine}.
\end{IEEEproof}

The extension to the asymmetric case is as follows. Let $K_A$ and $K_B$ respectively denote the number of requests for servers $A$ and $B$, and assume $K_A>K_B$ without loss of generality. Divide the $K=K_A+K_B$ requests (or users) into two groups: the first with $K_B$ requests for each server (symmetric demands) and the second with the remaining $K_A-K_B$ requests for server A. We construct effective pairs of length $t+1$ by appending requests from the second group to effective pairs from the first.

\begin{theorem}\label{t-2data1parity_asym}
If the requests are asymmetric, the ratio of unpaired messages is also bounded by $\Delta\leq\frac{1}{3}$. Specifically, if $K_A$ and $K_B$ respectively denote the number of requests for servers $A$ and $B$, assuming $K_A>K_B$ without loss of generality, the following normalized peak rate is achievable:
\begin{align}\label{e-pair}
R(K_A,K_B,t)&=\sum_{l=0}^{t+1}\binom{K_A-K_B}{l}R_T(2K_B,t-l),
%&\frac{1}{2}\binom{K_A-K_B}{t+1}\left/\binom{K}{t}\right.,
\end{align}
where $R_T$ is defined in Eq.~(\ref{e-rate}) and $K=K_A+K_B$.
\end{theorem}

\begin{IEEEproof}
From Lemma~\ref{l-pairing}, $R_T(2K_B,t-l)$ represents the peak rate after pairing all subsets of $t+1-l$ requests from the symmetric group. For each $l=0,1,\ldots,t+1$, we multiply $R_T(2K_B,t-l)$ by the number of possible completions with $l$ requests from the second group, to obtain the peak rate corresponding to subsets with $t+1-l$ requests from the first group and $l$ from the second. Adding them for all $l$ gives Eq.~(\ref{e-pair}).

Since $R_T(i,j)\leq \left(\frac{1}{2}+\frac{1}{6}\Delta\right)R_C(i,j)$ with $\Delta\leq\frac{1}{3}$ by Lemma~\ref{l-pairing}, and $\sum_{l=0}^{t+1}\binom{K_A-K_B}{l}R_C(2K_B,t-l)=R_C(K,t)$ by combinatorial equations, Eq.~(\ref{e-pair}) implies that $R(K_A,K_B,t)\leq \left(\frac{1}{2}+\frac{1}{6}\Delta\right)R_C(K,t)$ with $\Delta\leq\frac{1}{3}$ as defined in Lemma~\ref{l-unpaired}.
\end{IEEEproof}

\begin{corollary}
A peak rate of $\frac{5}{9}R_C(K,t)$ is achievable for a system with two data servers and a parity server.
\end{corollary}

\subsection{One parity and $L$ data servers}\label{ss-beyond}
The previous subsection has discussed the case with two data servers and one parity server, but the same algorithm can be extended to systems with more than two data servers. Intuitively, if there are $L$ data servers and one parity server, any message can be built by combining messages from any $L$ servers. A first approach could be distributing the $\binom{K}{t+1}$ messages in Maddah's scheme across the $L+1$ possible groups of $L$ servers, as proposed in subsection~\ref{ss-multiple}. Each server would then need to send a maximum of $\binom{K}{t+1}\cdot \frac{L}{L+1}$ messages. However, there is a more efficient way of fulfilling the requests based on the algorithms in subsections~\ref{ss-multiple}, \ref{ss-MoreFilesNoParity} and \ref{ss-algorithm}.

\begin{figure}
\centering
\includegraphics[width=0.18\textwidth,angle=270]{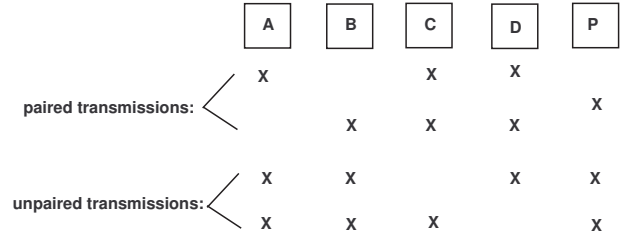}
\vspace{1em}
\caption{Pairing for $4$ data servers and one parity server system. $A,B,C,D$ are data servers and P represents the parity server. X means there is a message transmitted from the corresponding server.}
\label{fig:lservers}
\end{figure}

\begin{lemma}\label{l-transmit1}
Let $S_1=\{\mathbf{A}_{\pmb{\alpha}},\mathbf{B}_{\pmb{\beta}},\mathbf{A}_\ast,\mathbf{Y}\}$ and $S_2=\{\mathbf{A}_{\pmb{\alpha}},\mathbf{B}_{\pmb{\beta}},\mathbf{B}_\ast,\mathbf{Y'}\}$ be two user subsets, where $\mathbf{Y}$ and $\mathbf{Y'}$ are arbitrary lists of requests for servers $C$ through $L$ and the $\ast$ represent arbitrary (possibly empty) index sets. Then, $S_1$ and $S_2$ can be paired so that servers $A$, $B$ and $P$ require a single transmission to provide the same information as messages $m^{\mathbf{S_1}}$ and $m^{\mathbf{S_2}}$ in Maddah's single server scheme. The other data servers, $C$ through $L$, require a maximum of two transmissions, as shown in paired transmissions in Fig.~\ref{fig:lservers}.
\end{lemma}
\begin{IEEEproof}
The transmissions would proceed as follows:
\begin{enumerate}
\item Servers $C$ through $L$ each send two messages, to $S_1$ and $S_2$. For example, server $C$ would send $m_C^{S_1}$ and $m_C^{S_2}$, providing a desired segment to users requesting files from $C$ and the corresponding $C$-segments to those requesting other files.

\item Server $A$ sends\footnote{It would be enough for $A$ to send $m_A^{\{\mathbf{A}_\ast,\mathbf{A}_{\pmb{\alpha}},\mathbf{B}_{\pmb{\beta}}\}}$ instead of $m_A^{S_1}$, but we use the latter for the sake of simplicity. The same applies to the message from server $B$.} $m_A^{S_1}$, providing a desired segment to users requesting $\{\mathbf{A}_\ast,\mathbf{A}_{\pmb{\alpha}}\}$ and the corresponding undesired A-segments to those requesting $\mathbf{B}_{\pmb{\beta}}$.

\item Server $B$ sends $m_B^{S_2}$, providing a desired segment to users requesting $\{\mathbf{B}_{\pmb{\beta}},\mathbf{B}_\ast\}$ and the corresponding undesired B-segments to those requesting $\mathbf{A}_{\pmb{\alpha}}$.

\item Server $P$ sends $m_P^{\{\mathbf{A}_{\pmb{\alpha}},\mathbf{B}_{\pmb{\beta}}\}}$ to users requesting $\{\mathbf{A}_{\pmb{\alpha}},\mathbf{B}_{\pmb{\beta}}\}$. Using the undesired segments previously received, the users in $\{\mathbf{A}_{\pmb{\alpha}},\mathbf{B}_{\pmb{\beta}}\}$ can solve for the desired $A$ and $B$ segments.

\end{enumerate}
A simple comparison of the requested and received segments shows that these transmissions deliver the same information as messages $m^{\mathbf{S_1}}$ and $m^{\mathbf{S_2}}$ in Maddah's single server scheme.
\end{IEEEproof}

As an example, Table~\ref{t-beyondthree} shows the segments that each user gets in transmissions (1)-(4) when $S_1=\{A_1, A_2, B_1, C_1\}$ and $S_2=\{A_1, B_1, B_2, C_2\}$, respectively corresponding to segments $\{A_1^1,A_2^2,B_1^3,C_1^4 \}$ and $\{A_1^5,B_1^6,B_2^7,C_2^8 \}$.

\begin{table}
\centering\small
\begin{tabular}[htp]{|c|c|c|c|c|c|c|}
\hline
\tiny{Trans.$\setminus$}Req.& $A_1$ & $A_2$ & $B_1$ & $B_2$ & $C_1$ &$C_2$\\
\hline
(1)&$C_1^5$&&$C_1^3$&&$C_1^4$&$C_2^8$\\
\hline
(2)&$A_1^1$&$A_2^2$&$A_1^3$&&&\\
\hline
(3)&$B_1^5$&&$B_1^6$&$B_2^7$&&\\
\hline
(4)&$P_1^5$&&$P_1^3$&&&\\
\hline
in total&$A_1^1,A_1^5$&$A_2^2$&$B_1^3,B_1^6$&$B_2^7$&$C_1^4$&$C_2^8$\\
\hline
\end{tabular}
\caption{Segments received by each users in transmissions (1)-(4) from Lemma~\ref{l-transmit1}, where  $P_i^j=A_i^j\oplus B_i^j\oplus C_i^j$.}\label{t-beyondthree}
\end{table}

\begin{theorem}\label{t-rateLplus1}
The following normalized peak rate is achievable for a system with $L\geq 3$ data servers and one parity server:
\begin{equation}\label{e-oneparity}
R_P(K,t)=\frac{L-1}{L}R_C(K,t),
\end{equation}
where $R_C$ is defined in Eq.~(\ref{e-RC_binom}).
\end{theorem}
\begin{IEEEproof}
First we show that we can deliver $\frac{2}{L}\binom{K}{t+1}$ of the messages in Maddah's scheme using at most $\frac{1}{L}\binom{K}{t+1}$ transmissions from servers $A$, $B$ and $P$; and at most $\frac{2}{L}\binom{K}{t+1}$ transmissions from each of the other servers. This can be done by pairing the messages as shown in Lemma~\ref{l-transmit1}, if they include requests for $A$ or $B$, and by using the scheme in subsection~\ref{ss-MoreFilesNoParity}, if they do not.

Selecting these $\frac{2}{L}\binom{K}{t+1}$ messages can be done as follows: group messages by the number of segments that they have from servers $A$ or $B$. Within each group, we pair the messages as shown in Lemma~\ref{l-transmit1}. This is equivalent to pairing the $A$ and $B$ requests into effective pairs according to Theorem~\ref{t-2data1parity_asym} and considering all possible completions for each pair using requests for other servers.
Theorem~\ref{t-2data1parity_asym} showed that at least $\frac{2}{3}\geq \frac{2}{L}$ of the messages in each group can be paired. Messages which have no $A$ or $B$ segments can be transmitted as described in section~\ref{ss-MoreFilesNoParity}, without requiring any transmissions from servers $A$, $B$ or $P$.

The remaining $\frac{L-2}{L}\binom{K}{t+1}$ messages can be transmitted as described in subsection~\ref{ss-multiple}, distributing the savings evenly among servers $C$ through $L$. This requires $\frac{L-2}{L}\binom{K}{t+1}$ transmissions from servers $A$, $B$ and $P$; and $\frac{L-3}{L}\binom{K}{t+1}$ from each of the rest.

Each server then transmits a total of $\frac{L-1}{L}\binom{K}{t+1}$, hence the peak rate in Eq.~(\ref{e-oneparity}).
\end{IEEEproof}

Theorem~\ref{t-rateLplus1} provides a very loose bound for the peak rate in a system with one parity and $L$ data servers. In practice, there often exist alternative delivery schemes with significantly lower rates. For example, if all the users request files from the same server, that server should send half of the messages while all the other servers collaborate to deliver the other half. The rate would then be reduced to half of that in Maddah's scheme. Similarly, if $L>t+1$ and all the servers receive similar numbers of requests, the scheme in subsection~\ref{ss-MoreFilesNoParity} can provide significantly lower rates than Eq.~(\ref{e-oneparity}).

\subsection{Two parity and L data servers}\label{s-raid6}

In this section, we will extend our algorithm to a system with $L$ data and two linear parity servers operating in a higher order field instead of GF(2). The parity server $P$ stores the horizontal sum of all the files while the parity server $Q$ stores a different linear combination of the files BY ROW,  as shown in Table~\ref{t-sixservers}. It will be assumed that the servers form an MDS code. We will show that with a careful design of the delivery strategy, the peak rate can be reduced to almost half of that with Maddah's single server scheme.

\begin{table}
	\centering\small
	\begin{tabular}[h]{|c|c|}%
		\hline
		Server P & Server Q\\
		\hline
		$A_1 + B_1+\ldots+ L_1$ & $A_1+\kappa_BB_1+\ldots+\kappa_LL_1$\\
        $A_2 + B_2+\ldots+ L_2$ & $A_2+\kappa_BB_2+\ldots+\kappa_LL_2$\\
		$\vdots$ &$\vdots$ \\
        $A_r + B_r+\ldots+L_r$ & $A_r+\kappa_BB_r+\ldots+\kappa_LL_r$\\
		\hline
	\end{tabular}
	\caption{Files stored in parity servers in RAID-6}\label{t-sixservers}
\end{table}

\begin{lemma}\label{l-transmit2}
Let $S_1=\{\mathbf{A}_\ast,\mathbf{Y}\}$ and $S_2=\{\mathbf{B}_\ast,\mathbf{Y}\}$, where $\mathbf{Y}$ represents a common set of requests from any server. Then $S_1$ and $S_2$ can be paired so that a single transmission from each server fills the same requests as messages $m^{\mathbf{S_1}}$ and $m^{\mathbf{S_2}}$ in Eq.~(\ref{e-MNmessage}).
\end{lemma}
\begin{IEEEproof}
The transmission scheme shares the same pairing idea as the algorithm in subsection~\ref{ss-algorithm}. The transmissions are as follows:
\begin{enumerate}
\item Server A sends $m_A^{\mathbf{S_1}}$, providing a desired segment to users requesting its files and the corresponding undesired A-segments to others.

\item Server B sends $m_B^{\mathbf{S_2}}$, providing a desired segment to users requesting its files and the corresponding undesired B-segments to others.

\item Servers $C,D,\ldots,L$ each send a single message to $S_1\bigcap S_2=\{\mathbf{Y}\}$ with the following content for each user:
\begin{itemize}
	\item Users requesting files from server B received some undesired segments from server $A$. Servers $C,D,\ldots,L$ send them the matching ones so that the desired segments can be decoded using the parity in server $P$ later.
	\item The remaining users in $\mathbf{Y}$ will get the desired segment corresponding to $S_1$ when possible, otherwise they will get the undesired segment corresponding to $S_2$.
\end{itemize}
In other words, each server $C,\ldots,L$ will send segments corresponding to $S_1$ to users requesting its files or those from server $B$, and segments corresponding to $S_2$ to the rest.
At this point, all the users have satisfied their requests related to $S_1$, except those requesting files from server $B$, who satisfied their requests related to $S_2$ instead. Each user has also received $L-2$ undesired ``matched" segments\footnote{Users in $\mathbf{Y}$ requesting files from servers $A$ or $B$ received $L-1$ ``matched" segments instead of $L-2$, but we can ignore the extra one.}, corresponding to $S_1$ for those requesting files from server $B$ and corresponding to $S_2$ for the rest.

\item Finally, parity servers $P$ and $Q$ each transmit a message to $S_1\bigcap S_2=\{\mathbf{Y}\}$ with a combination of segments for each user (see Table~\ref{t-sixservers}). Those requesting files from server $B$ will get two combinations of the segments corresponding to $S_1$, while the rest will get two combinations of the segments corresponding to $S_2$. Since each user now has $L-2$ individual segments and two independent linear combinations of all $L$ segments, it can isolate the requested segment (as well all the ``matching" segments in other servers).
\end{enumerate}
A simple comparison of the requested and received segments shows that these transmissions deliver the same information as messages $m^{\mathbf{S_1}}$ and $m^{\mathbf{S_2}}$ in Maddah's single server scheme.
\end{IEEEproof}

As an example, Table~\ref{t-raid6} shows the delivered segments in transmissions
(1)-(4) if $m^{\mathbf{S_1}}=\{A_1^1,A_2^2,B_1^3,C_1^4,C_2^5\}$ and $m^{\mathbf{S_2}}=\{A_1^6,B_1^7,B_2^8,C_1^9,C_2^{10} \}$.

\begin{table}
	\centering\small
	\begin{tabular}[htp]{|c|c|c|c|c|c|c|}
		\hline
		\tiny{Trans.$\setminus$}Req.& $A_1$ & $A_2$ & $B_1$ & $B_2$ & $C_1$ &$C_2$\\
		\hline
		(1)&$A_1^1$&$A_2^2$&$A_1^3$&&$A_1^4$&$A_2^5$\\
		\hline
		(2)&$B_1^6$&&$B_1^7$&$B_2^8$&&\\
		\hline
		(3)&$C_1^6$&&$C_1^3$&&$C_1^9$&$C_2^{10}$\\
		\hline
		(4)&$P_1^6$&&$P_1^3$&&$P_1^4,Q_1^4$&$P_2^5,Q_2^5$\\
		\hline
		in total&$A_1^1,A_1^6$&$A_2^2$&$B_1^3,B_1^7$&$B_2^8$&$C_1^4,C_1^9$&$C_2^5,C_2^{10}$\\
		\hline
	\end{tabular}
	\caption{Segments users get in (1)-(4) transmissions (In order to simplify notation, denote $P_i^j=A_i^j+B_i^j+C_i^j$ and $Q_i^j=A_i^j+\kappa_BB_i^j+\kappa_CC_i^j$).}\label{t-raid6}
\end{table}
\begin{theorem}
For the $L$ data server and two parity server system, the following normalized peak rate is achievable:
\begin{equation}
R_Q(K,t)=\left(\frac{1}{2}+\frac{L-2}{2L+4}\Delta\right)R_C(K,t),
\end{equation}
where $\Delta\leq\frac{1}{3}$ is the pairing loss and $R_C$ is the rate of the single server Maddah's scheme in Eq.~(\ref{e-RC_binom}).
\end{theorem}

\begin{IEEEproof}
Group messages by the number of segments that they have from servers $A$ or $B$. Within each group, we pair the messages as shown in Lemma~\ref{l-transmit2}. If the number of requests from $A$ or $B$ is not zero, this is equivalent to pairing the $A$ and $B$ requests into effective pairs according to Theorem~\ref{t-2data1parity_asym} and considering all possible completions for each pair using requests for other servers. Theorem~\ref{t-2data1parity_asym} showed that at most $\frac{1}{3}$ of the messages in each group remains unpaired. % Thus the overall pairing loss $\Delta$ is also bounded by $\frac{1}{3}$.
For the messages which do not contain segments from $A$ or $B$ we repeat the same process with two other servers, with identical results: at most $\frac{1}{3}$ of them remain unpaired.

Each pair of messages can be delivered using a single transmission from each server, as shown in Lemma~\ref{l-transmit2}, hence paired messages contribute $\frac{1}{2}(1-\Delta)R_C(K,t)$ to the total rate, where  $\Delta$ denotes the ratio of unpaired messages. Unpaired messages are transmitted as described in section \ref{ss-multiple}, that is using $L$ out of the $L+2$ servers. Balancing this load among all the servers, they contribute $\frac{L}{L+2}\Delta R_C(K,t)$ to the total rate. Adding both contributions yields the rate above.
\end{IEEEproof}

\section{Scheme 2: Small cache}
\label{s-interference}
This section extends the interference elimination scheme in section~\ref{ss-interference} to a multi-server system. The interference elimination scheme is specially designed to reduce the peak rate when the cache size is small \cite{tian2016caching}. Unlike Maddah's scheme, which caches plain segments, the interference elimination scheme proposes caching linear combinations of them. That way each segment can be cached by more users, albeit with interference. This section will start with the system without parity in Table~\ref{t-servers_general}, showing that the transmission rate decreases as $\frac{1}{L}$ with the number of servers. Then it performs a similar analysis for the case with parity servers, which can be interpreted as an extension of the user's caches.

\begin{theorem}
In a system with $L$ data servers and parallel channels, the peak rate of the interference cancelling scheme can be reduced to $\frac{1}{L}$ of that in a single server system, \ie the following $(M,R)$ pair is achievable:
\begin{equation}\label{e-eliminationrate_lservers}
\left(\frac{t\left[(N-1)t+K-N\right]}{K(K-1)},\frac{N(K-t)}{LK}\right),\ t=0,1,\ldots,K.
\end{equation}
This holds regardless of whether each file is spread across servers (striping) or stored as a single block in one server.
\end{theorem}
\begin{IEEEproof}
Section~\ref{s-striping} showed that striping the files across $L$ servers reduces the peak rate of the interference cancelling scheme by $\frac{1}{L}$ compared with a single server system.

In contrast to Maddah's scheme, the interference cancelling scheme sends the same number of segments from each file, regardless of the users' requests. Moreover, each message consists of a combination of segments from a single file~\cite{tian2016caching}. Therefore, the same messages can be transmitted even if different files are stored in different servers. Each server will need to transmit a fraction $\frac{1}{L}$ of the messages, since it will be storing that same fraction of the files. The peak load can then be reduced to $\frac{1}{L}$ of that in Eq.~(\ref{e-eliminationrate}).
\end{IEEEproof}

If there are parity servers, we can further reduce the transmission rate by regarding them as an extension of the users' cache. Section \ref{ss-interference} explained that in the interference elimination algorithm~\cite{tian2016caching}, each user caches the parity symbols resulting from encoding a set of segments with a systematic MDS code $\mathcal{C}(P_0,P)$. It is possible to pick the code in such a way that some of these parity symbols can be found as combinations of the information stored in servers $P$ and $Q$. Then, instead of storing them in the user's cache, they are discarded. Those that are needed in the delivery phase will be transmitted by the parity servers.

For example, parity server $P$ stores the horizonal sum of the files, so it can transmit messages of the form:
\begin{equation}\nonumber
\sum_{i=1}^{N/L}\sum_{j=1}^{\binom{K-1}{t-1}}\lambda_{ij}\left(A_i^{\mathbf{s}_j}+B_i^{\mathbf{s}_j}\ldots+L_i^{\mathbf{s}_j}\right),
\end{equation}
with arbitrary coefficients $\lambda_{ij}$ for any user set $\mathbf{s}_j$. This corresponds to a linear combination of all the segments in Eq.~(\ref{e-segments_interf}). % and coefficients  are the coefficients to ensure the messages are linearly independent. These messages can be transmitted to this specific user to work as the linear combinations in the $\mathcal{C}(P_0,P)$ MDS code. Therefore, instead of storing linear components in the users, the parity server can transmit some of the components. We utilize the parity servers as shared common memory.
Similarly, parity server $Q$ can transmit some other linear combinations of the segments which can also work as components of an MDS code. %For example, in RAID-6 multi-server system with parity servers in Table~\ref{t-raid6}, both $P$ server and $Q$ server can transmit messages to each user which works as the cached parity symbols as shown in Fig.~\ref{fig:interference}. $M$ is the size of the cache at each user and $M''$ is the capacity of the messages transmitted by the parity servers. We call $M''$ as the ``common memory". So each user is equivalent to have a larger cache memory $M'=M+M''$ and the peak load is expected to be further reduced.
This effectively increases the size of the cache memories by $M'$ file units, corresponding to the amount of information that the parity servers can afford to send each user during the delivery phase.

\begin{theorem}\label{e-eliminationrateparity}
If there are $\eta$ parity servers and $K\geq N$, the following $(M,R)$ pairs are achievable for $t=0,1,\ldots,K$
\begin{equation}\nonumber
\left(\frac{t\left[(N-1)t+K-N\right]}{K(K-1)}-\eta\frac{N(K-t)}{LK^2},\frac{N(K-t)}{LK}\right).
\end{equation}
\end{theorem}
\begin{IEEEproof}
The information sent by the parity server is bounded by the peak rate of the data servers, \ie $\frac{N(K-t)}{LK}$ according to Eq.~(\ref{e-eliminationrate_lservers}). Assuming a worst case scenario, each transmission from a parity server will benefit a single user. Therefore, each parity server can effectively increase the cache of each user by $M'=\frac{N(K-t)}{LK^2}$.
\end{IEEEproof}

\begin{figure}
\centering
\includegraphics[width=1.2\textwidth]{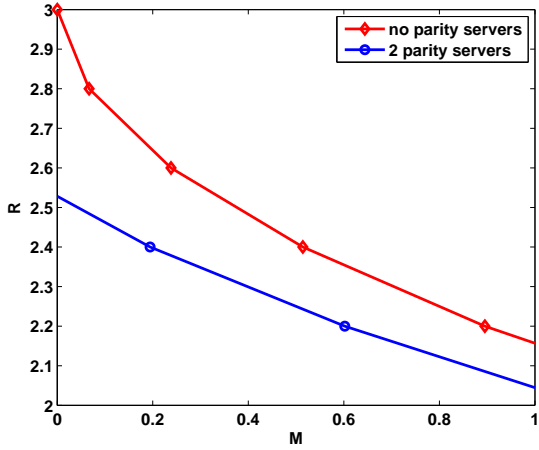}
\caption{Comparison of the performance between multi-server system without parity servers and the system with two parity servers.}
\label{fig:interference_compare}
\end{figure}

This memory sharing strategy provides significant improvement when the cache capacity is small. Fig.~\ref{fig:interference_compare} shows the performance for $K=15$ users  and $N=12$ files stored in $L=4$ data servers. When the cache size is small, the peak rate of the system with two parity servers is much lower than that without parity servers. As the cache grows the advantage of the system with parity servers becomes less clear.

The interference elimination scheme is specially designed for the case with less files than users ($N\leq K$) in the single server system. However, since the peak load is reduced by $\frac{1}{L}$ in a multi-server system, the interference elimination scheme might also have good performance when $N>K$ if $L$ is large. In order to apply the algorithm, we can just add $N-K$ dummy users with arbitrary requests. Then, we have the following corollary from Theorem~\ref{e-eliminationrateparity}:
\begin{corollary}
If there are $\eta$ parity servers and $K\leq N$, the following $(M,R)$ pairs are achievable:
\begin{equation}\nonumber
\left(\frac{t^2}{N}-\eta\frac{(N-t)}{LN},\frac{(N-t)}{L}\right),\ \ \ \ t=0,1,\ldots,N.
\end{equation}
\end{corollary}

\section{Simulations}\label{s-simulation}
This section compares all the schemes studied in this paper, for a system with $N=20$ files stored in $L=4$ data servers with $5$ files each. We show that striping has better performance than the schemes in sections \ref{s-ali} and \ref{s-interference} (Scheme~1 and Scheme~2, respectively) at the cost of network flexibility. If each file is stored as a single block in one server, Scheme~2 has better performance when the cache capacity is small while Scheme~1 is more suitable for the case where the cache capacity is large. The performances of Scheme~1 and Scheme~2 are summarized in Table~\ref{t-scheme1} and Table~\ref{t-scheme2}, respectively.
\begin{table}
	\centering
	\begin{tabular}[h]{|c|c|}%
		\hline
		server system & Normalized peak rate\\
		\hline
		single server & $R_C(K,t)=\binom{K}{t+1}/\binom{K}{t}$ \\
        \hline
		$L$ data $1$ parity& $\frac{L-1}{L}R_C(K,t)$ \\
        \hline
        $L$ data $2$ parity& $(\frac{1}{2}+\frac{L-2}{2L+4}\Delta)R_C(K,t)$ ($\Delta\leq\frac{1}{3}$) \\
		\hline
	\end{tabular}
	\caption{Normalized peak rate of Scheme~1.}\label{t-scheme1}
\end{table}
\begin{table}
	\centering
	\begin{tabular}[h]{|c|c|}%
		\hline
		server system & Normalized (M,R)\\
		\hline
		single server & $\left(\frac{t\left[(N-1)t+K-N\right]}{K(K-1)},\frac{N(K-t)}{K}\right)$ \\
        \hline
		$L$ data $\eta$ parity ($K\geq N$)& $\left(\frac{t\left[(N-1)t+K-N\right]}{K(K-1)}-\eta\frac{N(K-t)}{LK^2},\frac{N(K-t)}{LK}\right)$ \\
        \hline
        $L$ data $\eta$ parity ($K\leq N$)& $\left(\frac{t^2}{N}-\eta\frac{(N-t)}{LN},\frac{(N-t)}{L}\right)$ \\
		\hline
	\end{tabular}
	\caption{Normalized (M,R) pair of Scheme~2. ($\eta$ is the number of parity servers.)}\label{t-scheme2}
\end{table}

\begin{figure}
\centering
\includegraphics[width=1.2\textwidth]{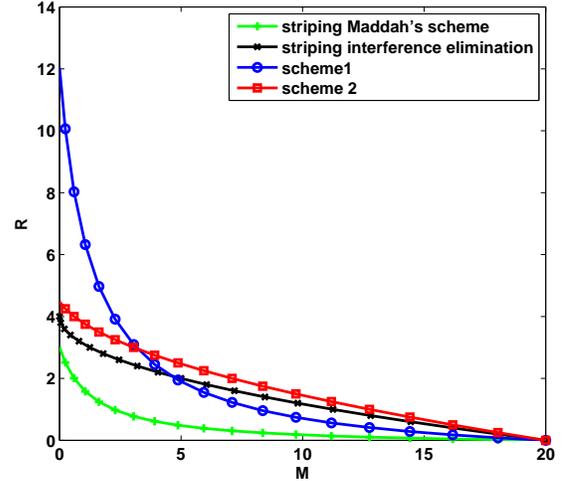}
\caption{Comparison between the performance between Scheme 1 and Scheme 2 in one parity server system when $N=20$ and $K=15$.}
\label{fig:compareraid4}
\end{figure}

\begin{figure}
\centering
\includegraphics[width=1.2\textwidth]{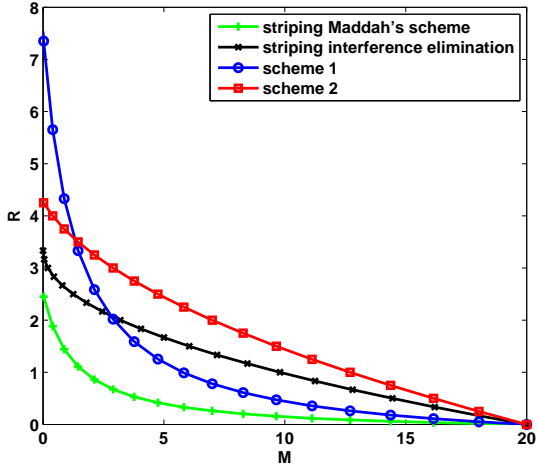}
\caption{Comparison between the performance between Scheme 1 and Scheme 2 in two parity server system when $N=20$ and $K=15$.}
\label{fig:compareNgreaterK}
\end{figure}

Fig.~\ref{fig:compareraid4} and Fig.~\ref{fig:compareNgreaterK} focus on the case with one and two parity servers, respectively. We assume that there are $K=15$ users, thus there are more files than users, with varying cache capacity. We observe that striping provides lower peak rates than storing whole files, as expected. Additionally, since $N>K$, the interference elimination scheme always has worse performance than Maddah's scheme when striping is used. Without striping, Scheme 2 provides lower peak rate than Scheme 1 when the cache capacity is small, and it is the other way around when the capacity is large.

\begin{figure}
\centering
\includegraphics[width=1.2\textwidth]{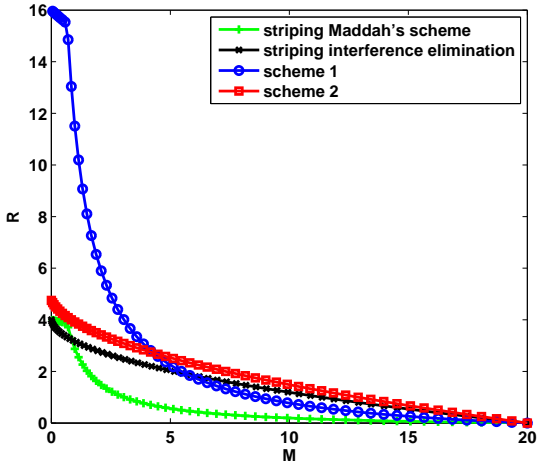}
\caption{Comparison between the performance between scheme 1 and scheme 2 in one parity server system when $N=20$ and $K=60$.}
\label{fig:compareraid4k}
\end{figure}

\begin{figure}
\centering
\includegraphics[width=1.2\textwidth]{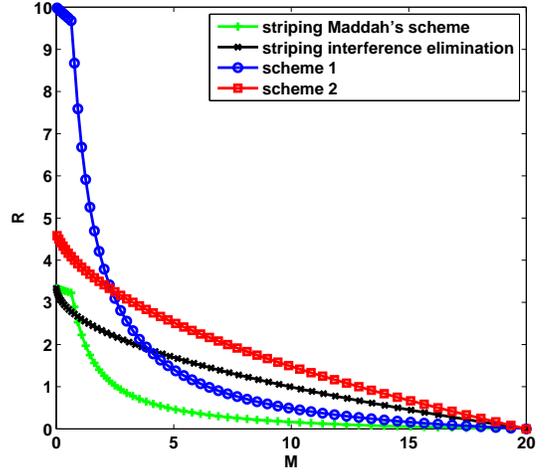}
\caption{Comparison between the performance between scheme 1 and scheme 2 in two parity server system when $N=20$ and $K=60$.}
\label{fig:compareKgreaterN}
\end{figure}

Then Fig.~\ref{fig:compareraid4k} and Fig.~\ref{fig:compareKgreaterN} compare the performance between Scheme~1 and Scheme~2 when there are more users $(K=60)$ than files for the one or two parity case, respectively. As shown in Fig.~\ref{fig:compareraid4k} and Fig.~\ref{fig:compareKgreaterN}, the striping has lower rate than storing whole files and when the cache capacity is very small, the striping interference elimination has better performance than striping Maddah's scheme. For Scheme~1 and Scheme~2, when the cache capacity is small, Scheme~2 provides lower peak rate, while when the cache capacity increases, Scheme~1 has better performance. Moreover, we notice that the curves intersect at a point with larger $M$ than they did in Fig.~\ref{fig:compareraid4} and Fig.~\ref{fig:compareNgreaterK}, which means that we are more prone to utilize Scheme~2 when there are more users than files.

\section{Conclusion}\label{s-conclude}
This paper proposes coded caching algorithms for reducing the peak data rate in multi-server systems with distributed storage and different levels of redundancy.
It shows that, by striping each file across multiple servers, the peak rate can be reduced proportionally to the number of servers. Then it addresses the case where each file is stored as a single block in one server and proposes different caching and delivery schemes depending on the size of the cache memories.

Distributed storage systems generally use MDS codes across the servers to protect the information against node failures. The coded caching schemes proposed in this paper are able to leverage that redundancy in creative ways to reduce the achievable traffic peak rate. The results for Scheme 1 and Scheme 2 are shown in Table~\ref{t-scheme1} and Table~\ref{t-scheme2} respectively.

In the future, we will study how this process can be generalized to other erasure codes, such as fractional repetition codes~\cite{el2010fractional}\cite{yu2014irregular} or other RAID-6~\cite{wang2014mdr} structures. We also plan to generalize our schemes to the case where files have different popularity, which will require designing erasure codes with different levels of protection for different files.

\appendix
In this appendix, we will elaborate on the pairing scheme in Lemma~\ref{l-pairing} from Section~\ref{ss-algorithm} , specially for the case with even $K$ and symmetric requests.

\begin{definition}\label{d-matching}
Let $\chi_A$ denote a set of messages (or, equivalently, subsets of $t+1$ users) to be sent by server A and $\chi_B$ denote a set of messages to be sent by server B. If there is an injective function providing each element in $\chi_A$ with an effective pair in $\chi_B$, we say that there is a \textbf{saturating matching} for $\chi_A$.
\end{definition}

In order to reduce the peak rate we want to separate all the messages to be transmitted (equivalently, subsets of $t+1$ users) into two groups $\chi_A$ and $\chi_B$ such that there are as many effective pairs as possible, as we shall see.

To better illustrate the allocation scheme, the problem of finding effective pairs is mapped to a graph problem. Let G be a finite bipartite graph with bipartite sets $\chi_A$ and $\chi_B$, where each message (or user subset) is represented as a vertex in the graph and edges connect effective pairs from $\chi_A$ and $\chi_B$. %$\chi_A=\{S_{A,1},S_{A,2},\ldots,S_{A,m}\}$ as defined in definition \ref{d-matching} and $S_{A,i}$ is represented as a vertex in the graph; same for $\chi_B$. If $\{S_{A,i},S_{B,j}\}$ forms an effective pair, we connect them by an edge.
The idea of our design is to allocate as many messages as possible to $\chi_A$, while guaranteeing the existence of a saturating matching for $\chi_A$ based on Hall's marriage Theorem \cite{hall1935representatives}.

\begin{theorem}\label{th-hall}
\emph{(Hall's Marriage Theorem \cite{hall1935representatives})} Let G be a finite bipartite graph with bipartite sets $\chi_A$ and $\chi_B$. For a set $\mathbf{u}$ of vertices in $\chi_A$, let $N_G(\mathbf{u})$ denote its neighbourhood in G, \ie the set of all vertices in $\chi_B$ adjacent to some element of $\mathbf{u}$. There is a matching that entirely covers $\chi_A$ if and only if
\begin{equation}\nonumber
|\mathbf{u}|\leq |N_G(\mathbf{u})|
\end{equation}
for every subset $\mathbf{u}$ of $\chi_A$.
\end{theorem}

\begin{corollary}\label{l-match}
If all vertices in $\chi_A$ have the same degree $d_A$ and all the vertices in $\chi_B$ have the same degree $d_B$ $(d_A\geq d_B)$, then there is a saturating matching for $\chi_A$.
\end{corollary}
\begin{IEEEproof}
 For any $\mathbf{u}\subseteq \chi_A$, all edges connected to $\mathbf{u}$ are also connected to $N_G(\mathbf{u})$, hence $|N_G(\mathbf{u})|\cdot d_B\geq |\mathbf{u}|\cdot d_A$. Since $d_A\geq d_B$, we know that $|\mathbf{u}|\leq |N_G(\mathbf{u})|$. According to Theorem~\ref{th-hall}, there is a saturating matching for $\chi_A$.
\end{IEEEproof}

In order to compute the peak rate in the worst case, we assume that all $K$ users request different files. Since each subset contains $t+1$ files, there are $\binom{K}{t+1}$ messages to allocate between $\chi_A$ and $\chi_B$. We classify these subsets according to the number of requests from server A: sets of \emph{type $w$} will have $w$ requests from server A and $t+1-w$ from server B. The following proposition states that the messages of the same type  are not able to pair with each other.

\begin{figure}
\centering
\includegraphics[width=0.15\textwidth,angle=270]{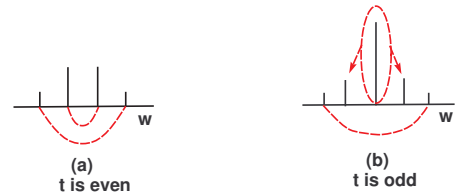}
\caption{Pairing illustration. $w$ is the number of files from server $A$ in a message.}
\label{fig:pair}
\end{figure}

When $t$ is even and the demands are symmetric, \emph{type $w$} sets and \emph{type $t+1-w$} sets form a symmetric bipartite graph, so there exists a saturating matching according to Corollary~\ref{l-match}. When $t$ is odd, \emph{type $(t+1)/2$} sets are paired with the union of \emph{type $(t-1)/2$} sets and \emph{type $(t+3)/2$} sets. Since the vertices in \emph{type $(t-1)/2$} sets and \emph{type $(t+3)/2$} sets are connected to the same number of vertices in \emph{type $(t+1)/2$} sets, this bipartite graph also fulfills the condition in Corollary~\ref{l-match}. Other sets are paired as in the case with $t$ even, that is, \emph{type $w$} sets are paired with \emph{type $t+1-w$} sets. These pairings are illustrated in Fig.\ref{fig:pair}.

When $t$ is even, there is a matching for every candidate file set, thus the peak rate is cut by half compared with the traditional single server scheme. When $t$ is odd, some vertices of types $(t-1)/2$, $(t+1)/2$, or $(t+3)/2$ could fail to be paired. Denote the ratio of unpaired messages when $t$ is odd by $\Delta$. Any two servers can collaborate to fulfill those requests, so the normalized overall peak rate $R_T$ with symmetric demands is given by:
\begin{displaymath}
R_T(K,t) = \left\{ \begin{array}{ll}
\frac{1}{2}R_C(K,t)&\textrm{if $t$ is even}\\
&\\
\left(\frac{1}{2}+\frac{1}{6}\Delta\right)R_C(K,t)&\textrm{if $t$ is odd,}
\end{array} \right.
\end{displaymath}

The pairing loss $\Delta$ is limited. The worst case occurs when there is a big difference between the number of vertices of type $(t+1)/2$ and the number of vertices of types $(t-1)/2$ or $(t+3)/2$. In both cases, the pairing loss $\Delta$ is bounded by $\frac{1}{3}$.

\bibliography{coded_b}

\begin{thebibliography}{10}
\providecommand{\url}[1]{#1}
\csname url@samestyle\endcsname
\providecommand{\newblock}{\relax}
\providecommand{\bibinfo}[2]{#2}
\providecommand{\BIBentrySTDinterwordspacing}{\spaceskip=0pt\relax}
\providecommand{\BIBentryALTinterwordstretchfactor}{4}
\providecommand{\BIBentryALTinterwordspacing}{\spaceskip=\fontdimen2\font plus
\BIBentryALTinterwordstretchfactor\fontdimen3\font minus
  \fontdimen4\font\relax}
\providecommand{\BIBforeignlanguage}[2]{{%
\expandafter\ifx\csname l@#1\endcsname\relax
\typeout{** WARNING: IEEEtran.bst: No hyphenation pattern has been}%
\typeout{** loaded for the language `#1'. Using the pattern for}%
\typeout{** the default language instead.}%
\else
\language=\csname l@#1\endcsname
\fi
#2}}
\providecommand{\BIBdecl}{\relax}
\BIBdecl

\bibitem{dimakis2010network}
A.~G. Dimakis, P.~Godfrey, Y.~Wu, M.~J. Wainwright, and K.~Ramchandran,
  ``Network coding for distributed storage systems,'' \emph{IEEE Trans. Inf.
  Theory}, vol.~56, no.~9, pp. 4539--4551, 2010.

\bibitem{ernvall2013capacity}
T.~Ernvall, S.~El~Rouayheb, C.~Hollanti, and H.~V. Poor, ``Capacity and
  security of heterogeneous distributed storage systems,'' \emph{IEEE J. Sel.
  Areas Commmun.}, vol.~31, no.~12, pp. 2701--2709, 2013.

\bibitem{corbett2004row}
P.~Corbett, B.~English, A.~Goel, T.~Grcanac, S.~Kleiman, J.~Leong, and
  S.~Sankar, ``Row-diagonal parity for double disk failure correction,'' in
  \emph{FAST-2004: 3rd Usenix Conference on File and Storage Technologies},
  2004.

\bibitem{plank2009raid}
J.~S. Plank, ``The {RAID}-6 liber8tion code,'' \emph{International Journal of
  High Performance Computing Applications}, 2009.

\bibitem{balasubramanian2014sap}
B.~Balasubramanian, T.~Lan, and M.~Chiang, ``{SAP}: Similarity-aware
  partitioning for efficient cloud storage,'' in \emph{IEEE INFOCOM 2014-IEEE
  Conference on Computer Communications}.\hskip 1em plus 0.5em minus
  0.4em\relax IEEE, 2014, pp. 592--600.

\bibitem{maddah2014fundamental}
M.~A. Maddah-Ali and U.~Niesen, ``Fundamental limits of caching,'' \emph{IEEE
  Trans. Inf. Theory}, vol.~60, no.~5, pp. 2856--2867, 2014.

\bibitem{ghasemi2015improved}
H.~Ghasemi and A.~Ramamoorthy, ``Improved lower bounds for coded caching,'' in
  \emph{IEEE Int. Symp. on Information Theory (ISIT)}.\hskip 1em plus 0.5em
  minus 0.4em\relax IEEE, 2015, pp. 1696--1700.

\bibitem{niesen2014coded}
U.~Niesen and M.~A. Maddah-Ali, ``Coded caching with nonuniform demands,'' in
  \emph{IEEE Conf. on Computer Communications Workshops (INFOCOM
  WKSHPS)}.\hskip 1em plus 0.5em minus 0.4em\relax IEEE, 2014, pp. 221--226.

\bibitem{zhang2015coded}
J.~Zhang, X.~Lin, C.-C. Wang, and X.~Wang, ``Coded caching for files with
  distinct file sizes,'' in \emph{IEEE Int. Symp. on Information Theory
  (ISIT)}.\hskip 1em plus 0.5em minus 0.4em\relax IEEE, 2015, pp. 1686--1690.

\bibitem{zhang2015code}
J.~Zhang, X.~Lin, and X.~Wang, ``Coded caching under arbitrary popularity
  distributions,'' in \emph{Information Theory and Applications Workshop (ITA),
  2015}.\hskip 1em plus 0.5em minus 0.4em\relax IEEE, 2015, pp. 98--107.

\bibitem{ji2014caching}
M.~Ji, A.~M. Tulino, J.~Llorca, and G.~Caire, ``Caching and coded multicasting:
  Multiple groupcast index coding,'' in \emph{IEEE Global Conf. on Signal and
  Information Processing (GlobalSIP)}.\hskip 1em plus 0.5em minus 0.4em\relax
  IEEE, 2014, pp. 881--885.

\bibitem{ji2015caching}
M.~Ji, A.~Tulino, J.~Llorca, and G.~Caire, ``Caching-aided coded multicasting
  with multiple random requests,'' in \emph{IEEE Information Theory Workshop
  (ITW)}.\hskip 1em plus 0.5em minus 0.4em\relax IEEE, 2015, pp. 1--5.

\bibitem{hachem2015effect}
J.~Hachem, N.~Karamchandani, and S.~Diggavi, ``Effect of number of users in
  multi-level coded caching,'' in \emph{IEEE Int. Symp. on Information Theory
  (ISIT)}.\hskip 1em plus 0.5em minus 0.4em\relax IEEE, 2015, pp. 1701--1705.

\bibitem{shariatpanahi2015multi}
S.~P. Shariatpanahi, S.~A. Motahari, and B.~H. Khalaj, ``Multi-server coded
  caching,'' \emph{arXiv preprint arXiv:1503.00265}, 2015.

\bibitem{tian2016caching}
C.~Tian and J.~Chen, ``Caching and delivery via interference elimination,''
  \emph{arXiv preprint arXiv:1604.08600}, 2016.

\bibitem{blom1984optimal}
R.~Blom, ``An optimal class of symmetric key generation systems,'' in
  \emph{Workshop on the Theory and Application of of Cryptographic
  Techniques}.\hskip 1em plus 0.5em minus 0.4em\relax Springer, 1984, pp.
  335--338.

\bibitem{suh2011exact}
C.~Suh and K.~Ramchandran, ``Exact-repair {MDS} code construction using
  interference alignment,'' \emph{IEEE Trans. Inf. Theory}, vol.~57, no.~3, pp.
  1425--1442, 2011.

\bibitem{el2010index}
S.~El~Rouayheb, A.~Sprintson, and C.~Georghiades, ``On the index coding problem
  and its relation to network coding and matroid theory,'' \emph{IEEE Trans.
  Inf. Theory}, vol.~56, no.~7, pp. 3187--3195, 2010.

\bibitem{bar2011index}
Z.~Bar-Yossef, Y.~Birk, T.~Jayram, and T.~Kol, ``Index coding with side
  information,'' \emph{IEEE Trans. Inf. Theory}, vol.~57, no.~3, pp.
  1479--1494, 2011.

\bibitem{chaudhry2008efficient}
M.~A.~R. Chaudhry and A.~Sprintson, ``Efficient algorithms for index coding,''
  in \emph{IEEE INFOCOM Workshops}.\hskip 1em plus 0.5em minus 0.4em\relax
  IEEE, 2008, pp. 1--4.

\bibitem{el2010fractional}
S.~El~Rouayheb and K.~Ramchandran, ``Fractional repetition codes for repair in
  distributed storage systems,'' in \emph{48th Annual Allerton Conference on
  Communication, Control, and Computing}.\hskip 1em plus 0.5em minus
  0.4em\relax IEEE, 2010, pp. 1510--1517.

\bibitem{yu2014irregular}
Q.~Yu, C.~W. Sung, and T.~H. Chan, ``Irregular fractional repetition code
  optimization for heterogeneous cloud storage,'' \emph{IEEE J. Sel. Areas
  Commun.}, vol.~32, no.~5, pp. 1048--1060, 2014.

\bibitem{huang2013pyramid}
C.~Huang, M.~Chen, and J.~Li, ``Pyramid codes: Flexible schemes to trade space
  for access efficiency in reliable data storage systems,'' \emph{ACM
  Transactions on Storage (TOS)}, vol.~9, no.~1, p.~3, 2013.

\bibitem{santos2000comparing}
J.~R. Santos, R.~R. Muntz, and B.~Ribeiro-Neto, \emph{Comparing random data
  allocation and data striping in multimedia servers}.\hskip 1em plus 0.5em
  minus 0.4em\relax ACM, 2000, vol.~28, no.~1.

\bibitem{wang2014mdr}
Y.~Wang, X.~Yin, and X.~Wang, ``{MDR} codes: A new class of {RAID}-6 codes with
  optimal rebuilding and encoding,'' \emph{IEEE J. Sel. Areas Commun.},
  vol.~32, no.~5, pp. 1008--1018, 2014.

\bibitem{hall1935representatives}
P.~Hall, ``On representatives of subsets,'' \emph{J. London Math. Soc},
  vol.~10, no.~1, pp. 26--30, 1935.

\end{thebibliography}
\end{document}